\documentclass[journal, 10pt]{IEEEtran}
\usepackage{cite}
\usepackage{soul}
\usepackage[pdftex]{graphicx}
\usepackage{amsmath}
\usepackage{amsfonts}
\usepackage{amssymb}
\usepackage{color}
\usepackage{float}
\usepackage{color}
\usepackage{epstopdf}
\usepackage{subfigure}
\usepackage[margin=0.5in]{geometry}
\usepackage{verbatim}
\usepackage{comment}
\usepackage{epsfig}
\usepackage{epstopdf}
\begin{document}

\title{Scaling Projections on Spin Transfer Torque Magnetic Tunnel Junctions}

\author{Debasis~Das, Ashwin~Tulapurkar and~Bhaskaran~Muralidharan

\thanks{The authors are with the Department of Electrical Engineering, IIT Bombay,
Mumbai 400076, India (e-mail: bm@ee.iitb.ac.in).}
}
\maketitle
\begin{abstract}
We investigate scaling of technologically relevant magnetic tunnel junction devices in the trilayer and pentalayer configurations by varying the cross-sectional area along the transverse direction using the non-equilibrium Green's function spin transport formalism. We study the geometry dependence by considering square and circular cross-sections. As the transverse dimension in each case reduces, we demonstrate that the transverse mode energy profile plays a major role in the resistance-area product. Both types of devices show constant tunnel magnetoresistance at larger cross-sectional areas but achieve ultra-high magnetoresistance at small cross-sectional areas, while maintaining low resistance-area products. We notice that although the critical switching voltage for switching the magnetization of the free layer nanomagnet in the trilayer case remains constant at larger areas, it needs more energy to switch at smaller areas. In the pentalayer case, we observe an oscillatory behavior at smaller areas as a result of double barrier tunneling. We also describe how switching characteristics of both kinds of devices are affected by the scaling.
\end{abstract}
\section{Introduction}
\IEEEPARstart{T}{h}ere is currently tremendous interest in spintronic devices with applications ranging from magnetic random access memories (MRAM), spin based logic devices \cite{Wang2005,Dery2007,Nikonov2011}  to spin torque nano-oscillators (STNO) \cite{Kaka2006,Houssameddine2007,Sharma2015,Kiselev2003,Katine2000} and magnetic field sensors \cite{VanDijken2005,Zeng2012}. Spin based memory devices have gained a lot of attention due to the non-volatile nature\cite{Parkin2003,Endoh2016},  low-power operation \cite{Deng2013,Augustine2012} and particularly, the possibility of writing data using spin currents \cite{Brataas2012,Chappert2007} via the phenomenon of spin-transfer torque (STT) \cite{slon,berger}.  There is hence a huge interest to further the spin-transfer torque magnetic random access memory (STT-MRAM) technology. \\
\indent Magnetic tunnel junctions (MTJs) are the building blocks of STT-MRAM technology, where information is stored in the form of the relative magnetic orientations of the constituent nanomagnets. Data stored in the MTJ is read out by sensing the resistance which depends on the relative orientation of the magnetization of the two ferromagnetic (FM) layers. Reading information from the MTJ depends on the relative orientation of magnetization directions between free and fixed FM layers, and is measured by tunnel magnetoresistance (TMR) defined as $TMR=(R_{AP}-R_P)/R_P$, where $R_{AP}$ and $R_P$ are the resistances of the device in the antiparallel and parallel configurations respectively. A large difference between these two resistances is required to distinguish two states, thus making high TMR as one of the essential criteria for effective MTJ operation. \\
\indent The basic requirement for high-density hard disk drives (HDDs) is to have (i) high TMR and (ii) low Resistance-Area (RA) product \cite{Maehara2011, Ikeda2005a} for the read head application. There is also considerable interest in pentalayer devices \cite{Chatterji,Sharma2016}, which feature resonant tunneling. These devices, which we term as resonant tunneling magnetic tunneling junctions (RTMTJ) offer ultrahigh TMR due to effective spin filtering. These aspects motivate us to explore the physics of MTJ and RTMTJ structures in the ultrasmall scale. Magnets with perpendicular magnetic anisotropy are a far better choice than the magnets with in-plane anisotropy, as they have a lower switching voltage with high thermal stability \cite{mangin2006current,Wang2013,zhang2012compact}. Perpendicular anisotropy can be achieved by Fe-rich CoFeB\cite{Yakata2009} or reducing the free layer CoFeB thickness less than 1.3nm\cite{Zeng2012,Ikeda2010}.\\
\indent In this paper, we investigate the scaling behavior of MTJ and RTMTJ devices by varying cross-sectional area from $10000\ nm^2$ to $25\ nm^2$ along the transverse direction.  Geometrical shape dependency of the devices is also studied. All the devices considered are simulated using spin-resolved non-equilibrium Green's function (NEGF) approach solved self-consistently with the Poisson's equation to capture the device physics accurately. In section II, we describe the device structure and the mathematical formulation of the background theory in detail. Following this, we explore the physics behind the scaling effect on TMR and RA products for MTJ devices in section III-A and RTMTJ devices in Section III-B. We also describe how the switching characteristics of both kinds of devices are affected with scaling in section III-C, and finally, in section IV, we propose a device where the properties under scaling may be utilized in real devices.

\section{Device Modeling}
\subsection{Theoretical Formulation}
\begin{figure}[h]
    \subfigure[]{\includegraphics[scale=0.25]{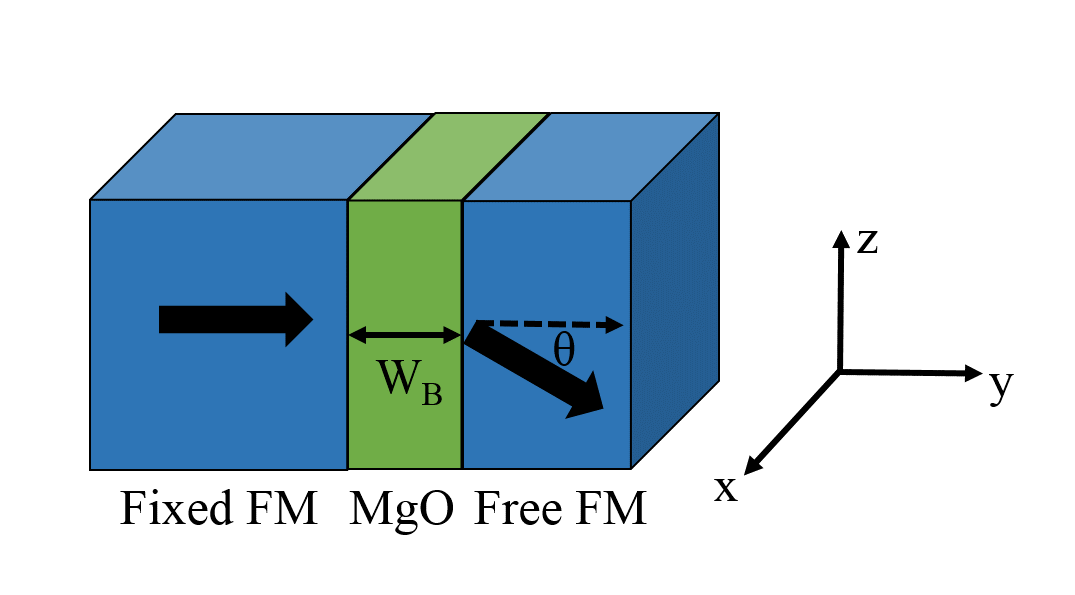}}
    \hfill
    \subfigure[]{\includegraphics[scale=0.25]{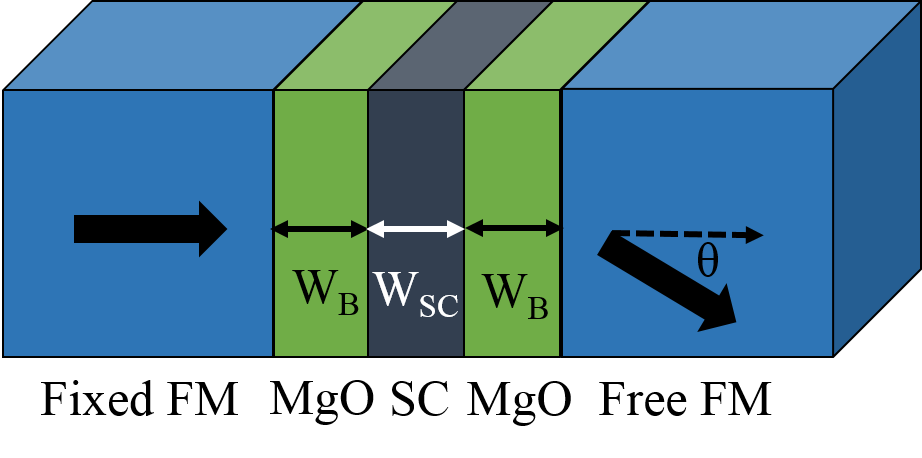}}
    \vfill
    \subfigure[]{\includegraphics[scale=0.25]{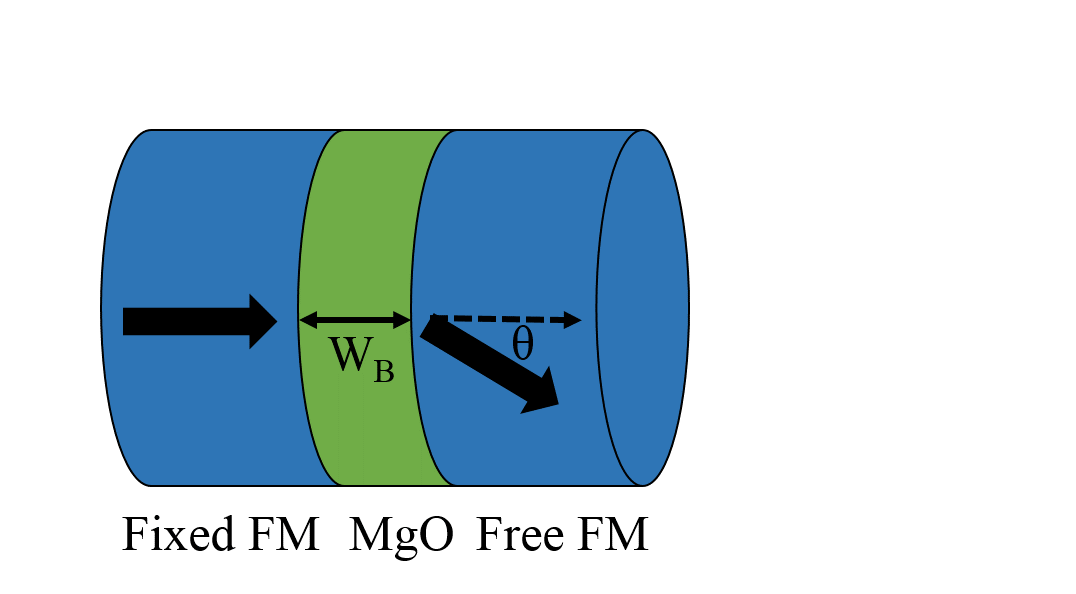}}
    \hfill
    \subfigure[]{\includegraphics[scale=0.25]{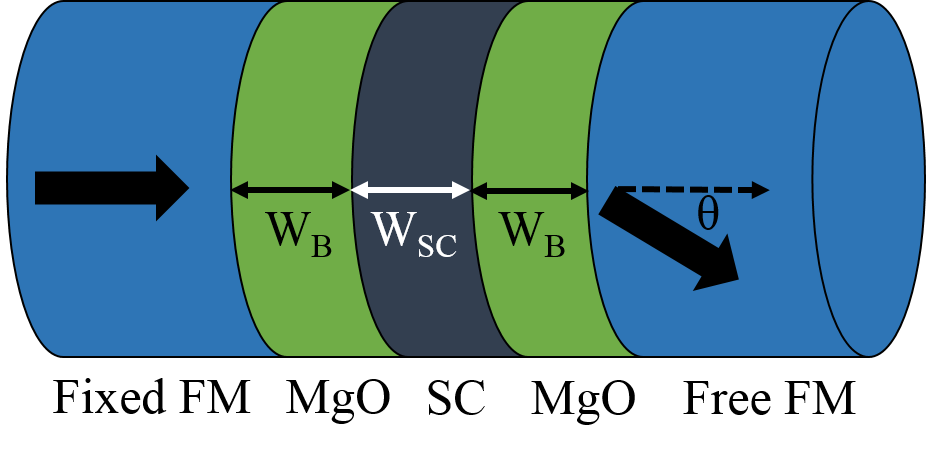}}
    \caption{Device structure: (a) Schematic of a trilayer MTJ having a square cross-section. An MgO  layer (green) is sandwiched between the fixed FM layer and the free FM layer. (b) Schematic of the pentalayer RTMTJ having a square cross-sectional transverse area. Between the two FM layers, a heterostructure of MgO (green)-Semiconductor (gray)-MgO (green) is sandwiched. (c) Schematic of a trilayer MTJ having circular cross-sectional transverse area. (d) Schematic of the pentalayer RTMTJ having a circular cross-sectional transverse area. $W_B$ and $W_{SC}$ denote width of the oxide and the semiconductor, respectively.}
    \label{Device_structure}
\end{figure}

In this section, we describe the device structures as shown in Fig. \ref{Device_structure} and the simulation procedure using the NEGF formalism in detail \cite{datta2,Yanik2007, Datta2012, Datta2010}. In a trilayer MTJ a thin MgO layer is sandwiched between a fixed and a free FM layer, whereas in RTMTJ a MgO-Semiconductor(SC)-MgO heterostructure is sandwiched between the fixed and the free FM layers. The FM layers are magnets with perpendicular anisotropy as it is technologically relevant. Here for all structures  
it is assumed that electronic transport occurs along the $\hat{y}$ direction, whereas the magnetization of the FM layers are in the $\hat{y}-\hat{z}$ plane. The magnetization of the fixed layer FM is assumed to be along $\hat{y}$ direction, whereas the magnetization of free layer FM is confined in the $\hat{y}-\hat{z}$ plane making an angle $\theta$ with the $\hat{y}$ direction. As the two magnetization directions are not collinear,  we need to perform a basis transformation\cite{Yanik2007} while calculating currents across the device. \\ 
\indent Owing to the length scales along the transport direction, we assume ballistic transport, so that the use of uncoupled mode space approach \cite{Venu} is justified. For investigating the scaling effects, we vary the transverse area in all devices as shown in Fig. \ref{Device_structure} from 25 $nm^2$ to 10000 $nm^2$. For each structure, we evaluate the transverse modes via the two-dimensional (2D) Schr\"{o}dinger equation described by the tight binding Hamiltonian of the cross-section and we assume that wave vector $\bf{k}$ is conserved for each transverse mode. The total charge current is evaluated by solving the one-dimensional (1D) NEGF formalism for each mode and finally summed over all the transverse modes. For each mode, we can assume the device as a 1D array of atoms. For NEGF we start with the energy resolved Green's function matrix $\left[ G(E)\right]$, calculated from the device Hamiltonian matrix $\left[ H\right] $ given by 
\begin{align}
	\left[ G(E)\right] &=\left[ EI-H-\Sigma_L-\Sigma_R\right]^{-1}\label{G_eq}\\
	\left[ H\right] &=\left[ H_0\right] +\left[ U\right],  \label{H_eq}
\end{align} 
where $\left[ H_0\right]$ matrix is calculated from the parameterized effective mass tight-binding approach \cite{Datta2012}, $\left[ U\right]$ is Coulomb charging matrix in real space, and $\left[ I\right]$ is the unitary matrix whose dimension is the same as the number of lattice points considered along the transport direction. Here, $\left[ \varSigma_L\right]$ and $\left[ \varSigma_R\right]$ represent self-energy matrices of the fixed and free FM layers respectively\cite{Sharma2016},\cite{datta2}. To ensure the Hamiltonian matrix to be hermitian we included the effect of an interface atom in the Hamiltonian which accounts for the varying parameters on both sides of it \cite{datta2}. An earlier work \cite{Neophytou2008}, established that the effective mass increases if the dimension of the device is reduced below 5 nm.\\
\indent For the trilayer MTJ, it can be assumed that there will be a linear potential drop across the oxide layer, whereas for the RTMTJ,  due to its multilayer nature, we capture the potential drop inside the device accurately via a self-consistent NEGF-Poisson solution.  Thus Coulomb charging matrix, $\left[ U\right]$, is calculated by solving following two equations self-consistently along the transport direction $y$ given by 
\begin{align}
	\frac{d}{dy}\left(\epsilon_r(y)\frac{d}{dy}U(y)\right)=-\frac{q^2}{\epsilon_0}n(y)\label{U_eq}\\
	n(y)=\frac{1}{2\pi A\ a_0} \sum_{k_x,k_z} G^n(y;k_x,k_z), \label{n_eq}
\end{align}
where, A is the cross-sectional area, $a_0$ represents the interatomic spacing and $G^n(y;k_x,k_z)=G^n(y,y,k_x,k_x,k_z,k_z)$ is a diagonal element of the electron correlation matrix given by
\begin{equation}
	[G^n]=\int dE\ \left[ [A_L(E)]f_L(E)+[A_R(E)]f_R(E)\right].
\end{equation}
Here, $[A_{L,R}(E)]=G(E)\Gamma_{L,R}(E)G^{\dagger}(E)$ is the spectral function, $\Gamma_{L,R}(E)=i([ \Sigma_{L,R}(E)] - [ \Sigma_{L,R}^{\dagger}(E)])$ represents the spin-dependent broadening matrices of the left(L) and the right(R) contact respectively, and $f_L(E)$ and $f_R(E)$ represent the Fermi-Dirac distribution of left and right contacts respectively. While solving Eq. (\ref{U_eq}) with Eq.(\ref{n_eq}) self-consistently, the boundary condition for $U$ is assumed to be $U_L=-qV/2$ and $U_R=qV/2$ at the left and the right contact regions, where $V$ is the applied voltage across the device. Once the Coulomb charging matrix $\left[ U\right] $ has been calculated, charge and spin currents across the device can be evaluated using the charge current operator $\tilde{I}_{op}$ between two neighboring lattice points $i$ and $i+1$ given by \cite{Datta2012} \\
\begin{equation}\label{Iop_eq}
	\tilde{I}_{op_{i,i+1}}=\frac{i}{\hbar}\left( H_{i,i+1}G^n_{i+1,i}-G^{n^{\dagger}}_{i,i+1}H^{\dagger}_{i+1,i}\right),
\end{equation}
where, $H$ and $G^n$ are the $2 \times 2$ Hamiltonian and correlation matrices of the system in spin space. The current operator $\tilde{I}_{op_{i,i+1}}$ is also in the spin basis represented by a $2\times 2$ matrix. The charge current $I$ and the spin current $\overrightarrow{I}_S$ from Eq. (\ref{Iop_eq}) are calculated as
\begin{align}
	I&=q\int dE\ Real[Trace(\tilde{I}_{op_{i,i+1}})] \label{Ic}\\
	\overrightarrow{I}_S&=q\int dE\ Real[Trace(\overrightarrow{\sigma}\tilde{I}_{op_{i,i+1}})].
\end{align}
Here, $q$ is electronic charge, $\hbar$ is the reduced Planck constant and $\overrightarrow{\sigma}$ is the Pauli spin vector.\\

\indent The magnetization dynamics of the free layer FM is described by the Landau-Lifshitz-Gilbert-Slonczewski (LLGS) equation given by \cite{dattaLN}
\begin{equation}\label{LLGS}
	(1+\alpha^2)\frac{\partial \hat{m}}{\partial t}=-\gamma \hat{m}\times \overrightarrow{H}_{eff}-\alpha \gamma(\hat{m}\times \hat{m}\times \overrightarrow{H}_{eff})-\overrightarrow{\tau}_{spin}
\end{equation}
where, $\alpha$ is the Gilbert damping parameter, $\hat{m}$ is the unit vector along the magnetization direction of the free FM layer, $\gamma$ is the gyromagnetic ratio and $\overrightarrow{H}_{eff}$ is the effective magnetic field given by
\begin{equation}\label{H_eff}
	\overrightarrow{H}_{eff}=\overrightarrow{H}_{ext}+\left( H_k-H_d\right) m_y \hat{y},
\end{equation}
 where $\overrightarrow{H}_{ext}$ is the externally applied magnetic field, $H_k=2K_{u2}/M_S$ is the anisotropy field, $H_d=4\pi M_S$ is the demagnetization field. $M_S$ is the saturation magnetization of the free layer magnet, and $K_{u2}$ is uni-axial anisotropy constant. The quantity $\overrightarrow{\tau}_{spin}$ in Eq. (\ref{LLGS}) is the spin torque exerted on the free layer FM, given by
\begin{equation}
	\overrightarrow{\tau}_{spin}=\frac{\gamma \hbar}{2qM_SV}\left[ \left( \hat{m}\times \hat{m}\times \overrightarrow{I}_S\right)-\alpha\left(  \hat{m}\times \overrightarrow{I}_S\right) \right],
\end{equation}

where, V is the volume of the free magnet. 

\subsection{Device Parameters}
\begin{figure}[h]
	\begin{center}
		\subfigure[]{\includegraphics[scale=0.16]{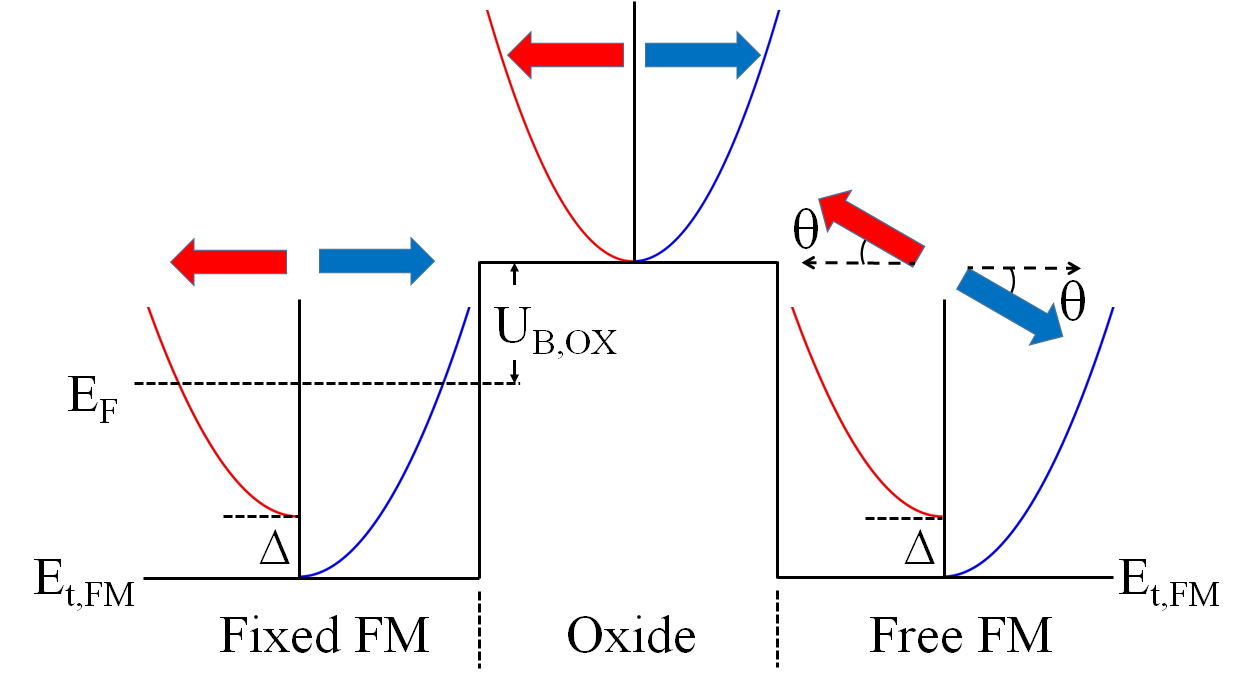}}
		\hfill
		\subfigure[]{\includegraphics[scale=0.16]{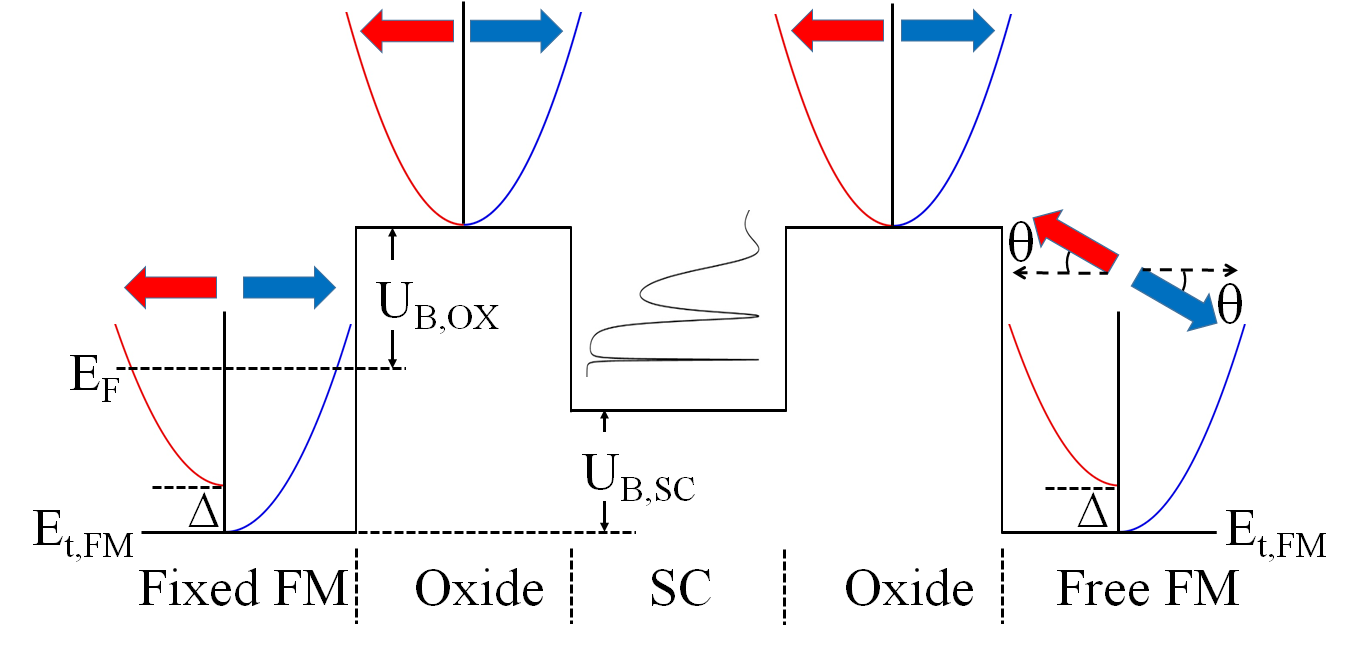}}
	\end{center}
	\caption{(a) Energy band diagram of a trilayer MTJ for a single transverse mode. The exchange splitting is $\Delta$ for the FM contact, $U_{B,OX}$ is the barrier height of the oxide layer above Fermi energy $E_F$. (b) Energy band diagram of a pentalayer RTMTJ for a single transverse mode. $U_{B,SC}$ is the energy difference between the bottom of the conduction band of the semiconductor and the FM. A sample transmission spectrum of the device with the resonant peaks is shown in the inset.}
	\label{device_band}
\end{figure}

For all the device structures, we used CoFeB as the FM for both the fixed and free layers with a Fermi energy $E_F$=2.25 eV and the exchange splitting $\Delta$=2.15 eV. The barrier height  from the Fermi level as shown in Fig. \ref{device_band} and the effective mass of MgO are considered as $U_{B,OX}$=0.76 eV and $m_{OX}$=0.18$m_e$ respectively, where $m_e$ is the free electron mass. Width of the oxide layer for all structures is assumed to be $W_B$=1 nm. For the RTMTJ structure, the width of the semiconductor is taken to be $W_{SC}$=1 nm with an effective mass $m_{SC}$=0.38$m_e$ and the well width $U_{B,SC}$=-0.45 eV. For simplicity, we used tight binding parameters reported in a related earlier work\cite{Sharma2016}. While calculating the TMR of the device, a small voltage difference between the two FM layers is kept constant for all the devices so that the magnetization of the free layer does not get affected by the spin transfer torque carried by the itinerant spins. 

\section{Results and Discussion}
\subsection{Trilayer MTJ}
\begin{figure}[h]
 	\subfigure[]{\includegraphics[scale=0.28]{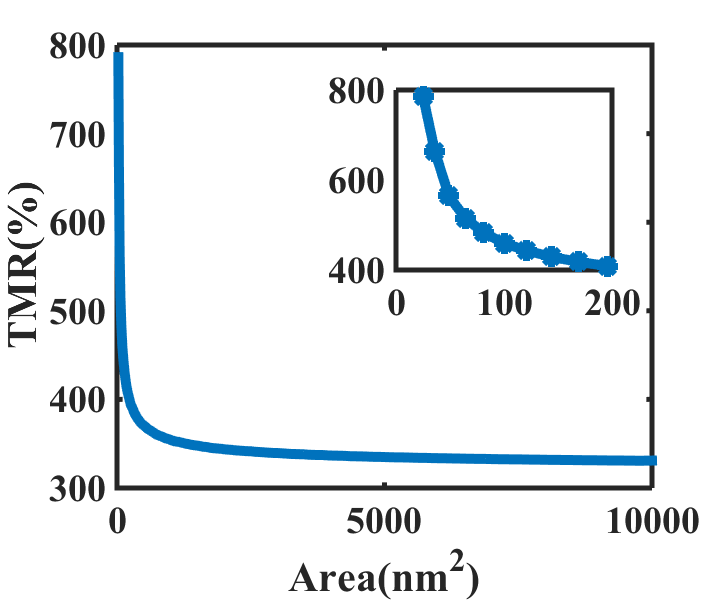}}
 	\hfill
 	\subfigure[]{\includegraphics[scale=0.28]{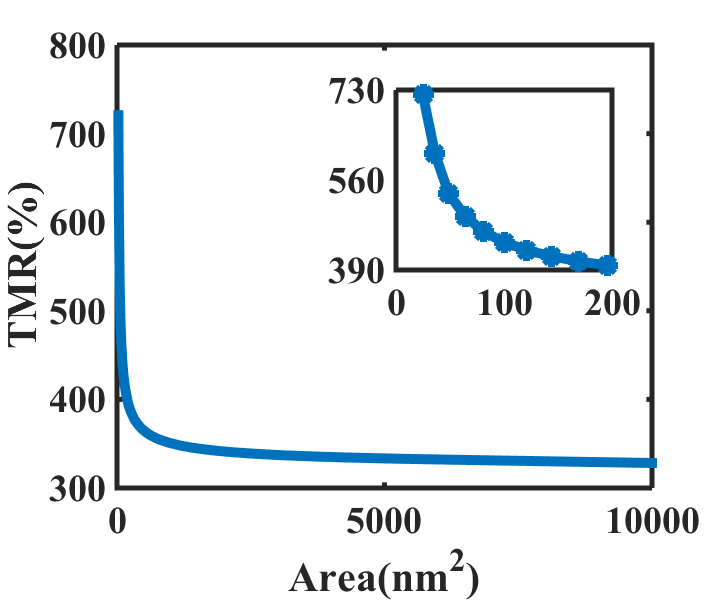}}

 	\subfigure[]{\includegraphics[scale=0.28]{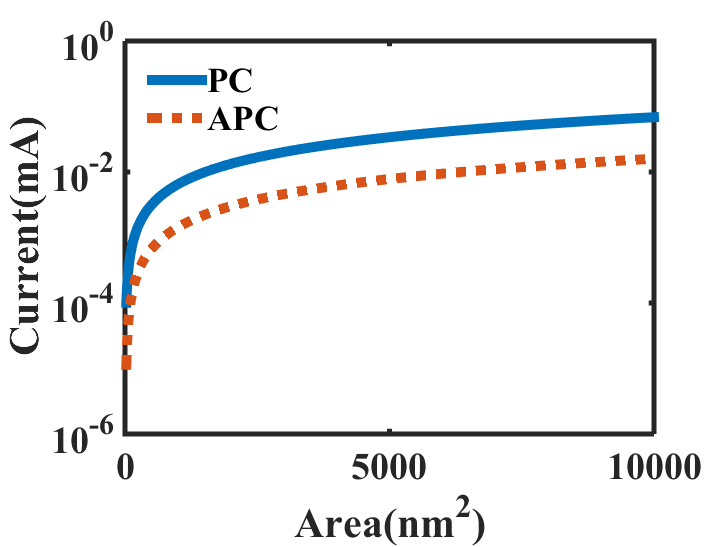}}
 	\hfill
 	\subfigure[]{\includegraphics[scale=0.28]{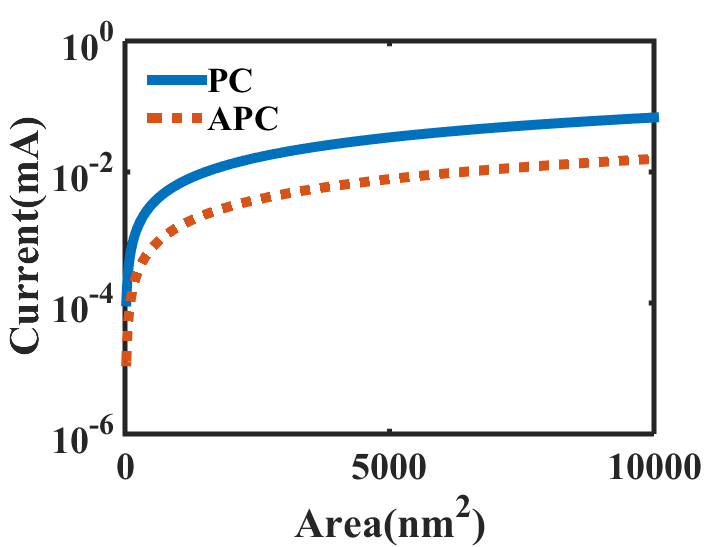}}
 	 	
 	\caption{TMR and charge current profile for the trilayer MTJ. Variation of TMR with cross-sectional area for (a) Square cross-section and (b) Circular cross-section. Insets show variation of TMR for the smaller areas. The `*' symbols in the inset figures denote data points. Variations in charge currents with cross-sectional areas in the PC and APC for (c) Square cross-section and (d) Circular cross-section. }
 	
 	\label{Tri_TMR_current_area}
 \end{figure}

\begin{figure}[h]
	\subfigure[]{\includegraphics[scale=0.28]{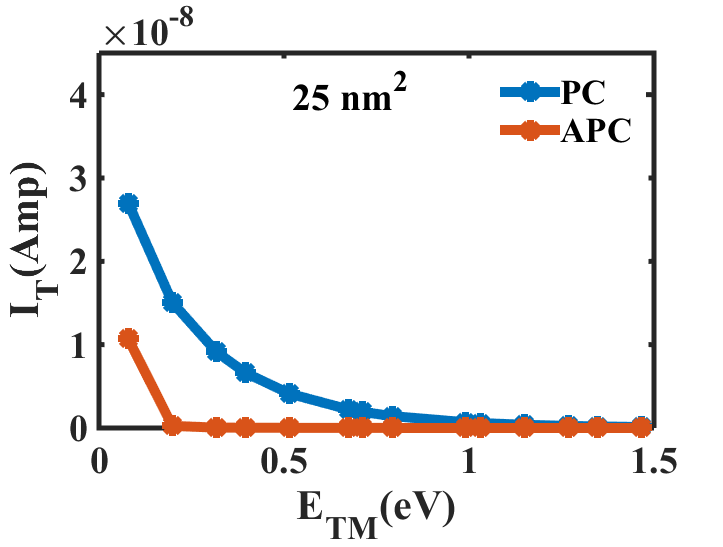}}
	\hfill
	\subfigure[]{\includegraphics[scale=0.28]{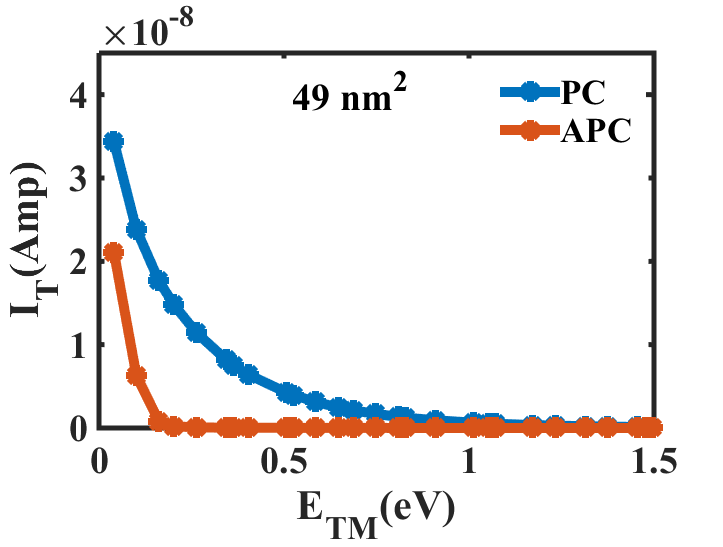}}
	\vfill
	\subfigure[]{\includegraphics[scale=0.28]{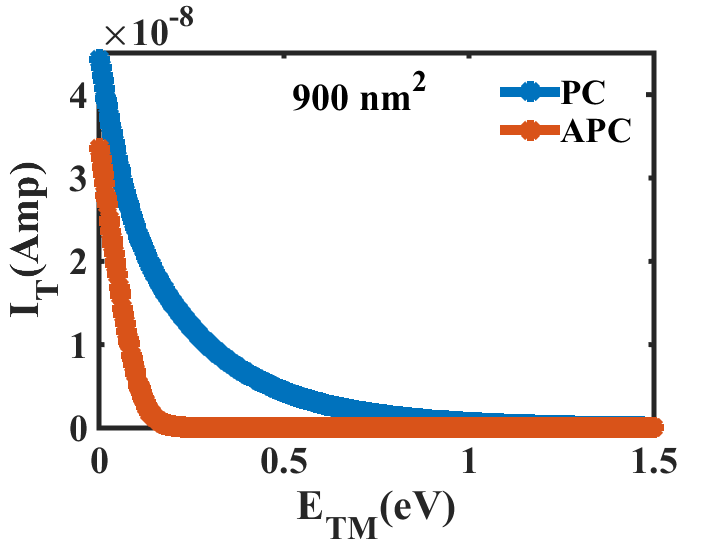}}
	\hfill
	\subfigure[]{\includegraphics[scale=0.28]{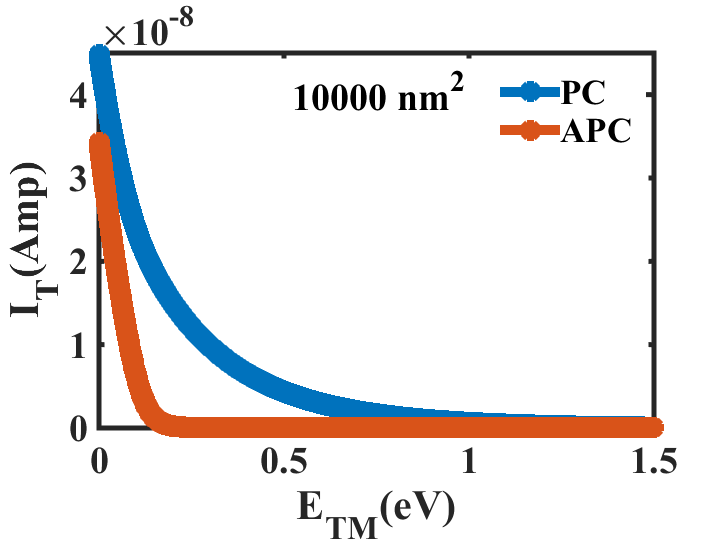}}
	\vfill
	\subfigure[]{\includegraphics[scale=0.28]{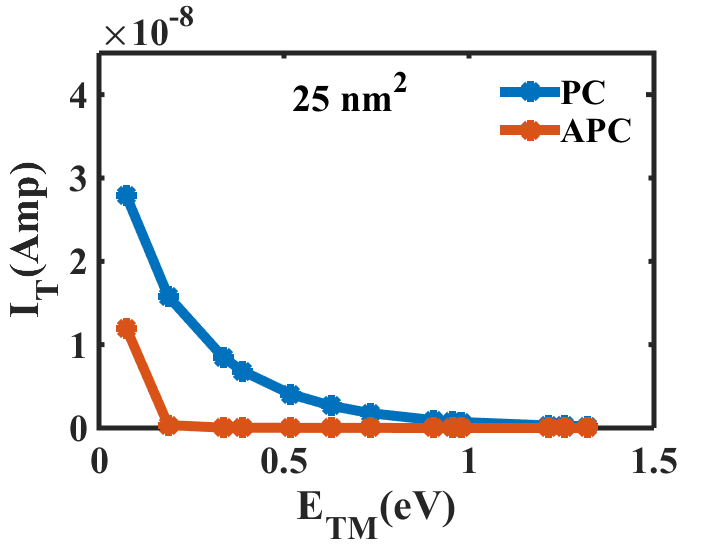}}
	\hfill
	\subfigure[]{\includegraphics[scale=0.28]{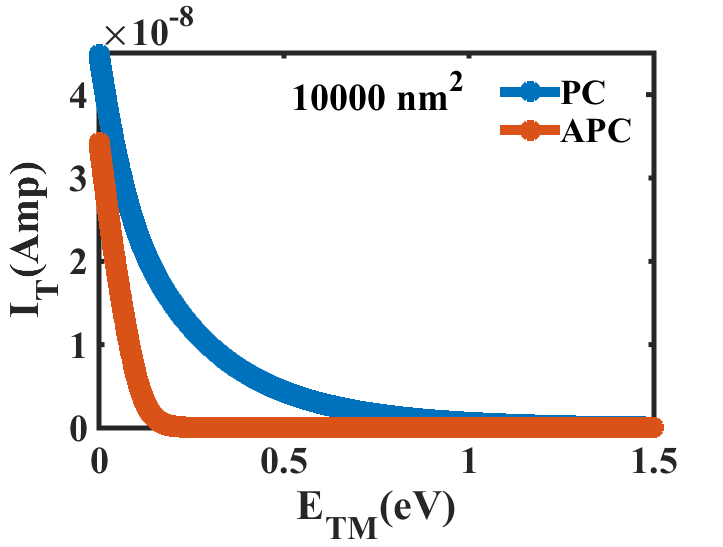}}

	\caption{Transverse mode profiles for a trilayer MTJ in PC and APC. Variations in transverse mode currents with transverse mode energy for a square cross-sectional area of (a) $25\ nm^2$, (b) $49\ nm^2$, (c) $900\ nm^2$, (d) $10000\ nm^2$ and for a circular cross-sectional area of (e) $25\ nm^2$ and (f) $10000\ nm^2$. }
	\label{trilayer_mode_profie}
\end{figure}

In this section, we describe the impacts of scaling on trilayer MTJ devices with square and circular cross-sections. First, we discuss how the TMR is affected by the variation of the area for both types of devices as shown in Fig. \ref{Tri_TMR_current_area}. The TMR is calculated as 
\begin{equation}\label{TMR_eq}
TMR=\frac{R_{AP}-R_P}{R_P}=\frac{I_P-I_{AP}}{I_{AP}}=\frac{I_P}{I_{AP}}-1 ,
\end{equation}
 where, $R_P$, $R_{AP}$ are resistance and $I_P$ and $I_{AP}$ are charge current of the device in parallel configuration (PC) and antiparallel configuration (APC) respectively.
 For both 
 characteristics, we see that for a larger cross-sectional area, the TMR remains constant at a value of 330\%, but it shoots up rapidly as the area decreases to a smaller value. To facilitate a deeper analysis, we highlight the sharp-increasing zone, as shown in the inset of Fig. \ref{Tri_TMR_current_area} (a) and (b) for both the square and circular cross-sections respectively. These plots reveal that instead of shooting up rapidly, the TMR increases at a finite rate as the cross-sectional area decreases. 
\noindent In our simulations, the highest TMR that we obtain for the MTJ at the smallest area is 786\% and 721\% for square and circular cross-sections respectively. This characteristic can be understood from the current dependency on the cross-sectional area for both PC and APC as shown in Fig. \ref{Tri_TMR_current_area} (c) and (d). For smaller cross-sectional areas, the currents for both PC and APC sharply increase whereas it increases at a constant rate as the area increases. As discussed earlier, the mode space approach is used to calculate currents, due to which the areal variation of currents can be explained using the transverse mode current profile for different areas as shown in Fig. \ref{trilayer_mode_profie}, where the energy profile of the transverse mode currents in the device is plotted.
\noindent  Figure \ref{trilayer_mode_profie} (a)-(d) shows the mode profile for the square cross-section, whereas (e)-(f) shows that for the circular cross-section for various area. As it is known, tighter the confinement, smaller the number of modes with larger energy spacing. Hence, it is expected that the mode energies will be closer if the area increases and becomes quasi-continuous with increasing areas. The smallest area, 25 $nm^2$, that we considered for both square and circular cross-sections, allows a few discrete modes to conduct with large energy spacing as shown in Fig. \ref{trilayer_mode_profie} (a) and (e). For the APC, very few modes carry the current in comparison to the PC, which creates a large difference between charge currents in the two configurations, thus creating a large TMR. For the simulation of trilayer MTJ, we considered the cut-off energy for the transverse modes as 1.5 eV, whereas practically, mode current contribution is up to 0.2 eV and 1.2 eV for APC and PC respectively.

\noindent As the area is increased slightly higher than the lowest one, the number of modes increases, thus decreasing the energy spacing between consecutive modes for both PC and APC, as shown in Fig. \ref{trilayer_mode_profie} (b). From this figure it is also seen that APC mode profile is much steeper than the PC one, so due to the increase of transverse modes with increasing area, current for both configuration increase but the rate of increase of current in APC is greater than PC, thus decreases TMR. This explains the phenomenon of decreasing TMR with an increase in cross-sectional area as shown in the inset of Fig. \ref{Tri_TMR_current_area} (a) and (b). When the area is increased further, energy spacing between consecutive modes becomes smaller and the mode profile looks like a quasi-continuous curve for both PC and APC as shown in Fig. \ref{trilayer_mode_profie} (c)-(d). In the range of larger areas, upon increasing the area, the number of modes increases and due to their quasi-continuous nature, the current increases at a constant rate for both configurations, thus making the TMR to be almost constant.\\
\begin{figure}[h]
	\subfigure[]{\includegraphics[scale=0.28]{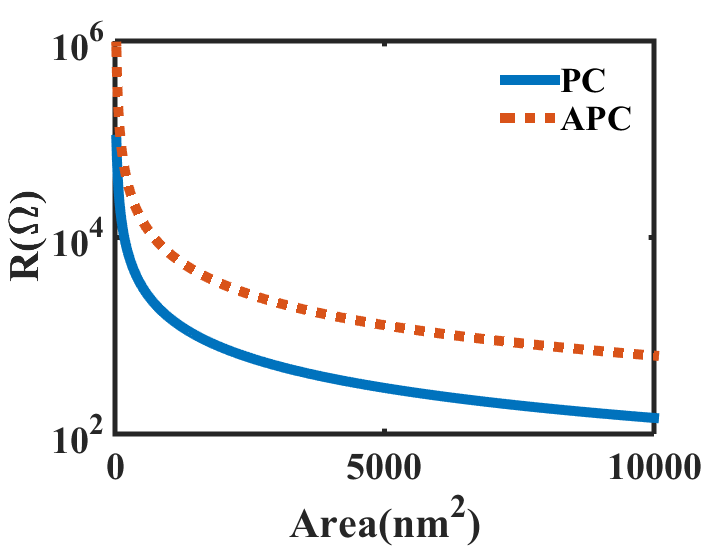}}
	\hfill
	\subfigure[]{\includegraphics[scale=0.28]{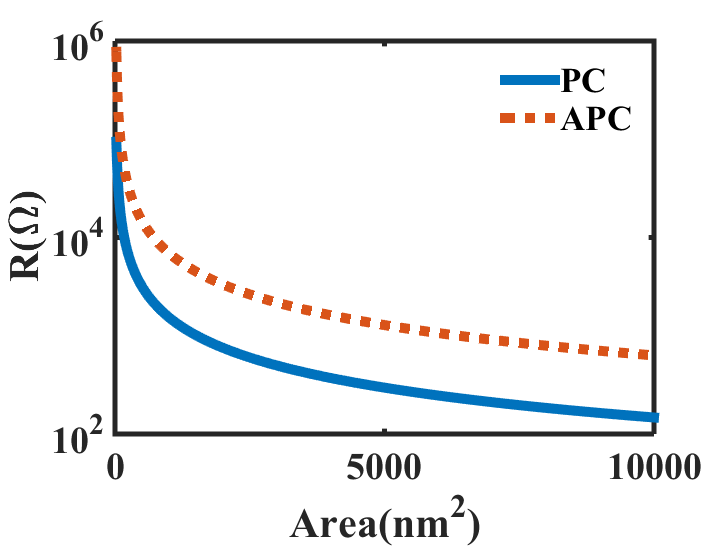}}
	
	\subfigure[]{\includegraphics[scale=0.28]{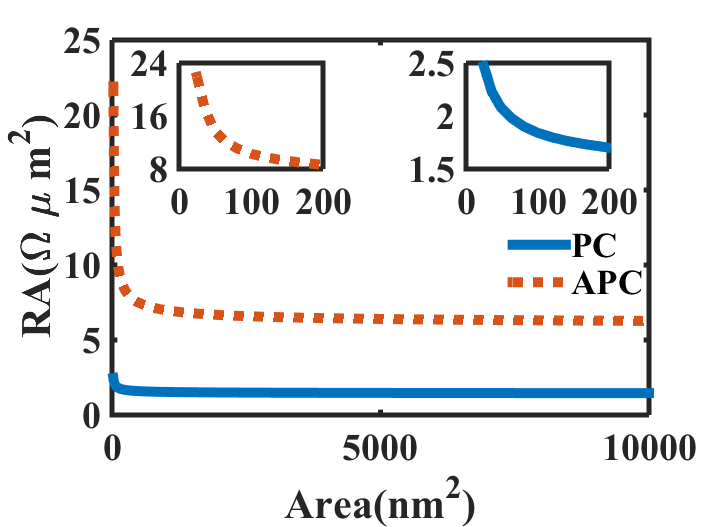}}
	\hfill
	\subfigure[]{\includegraphics[scale=0.28]{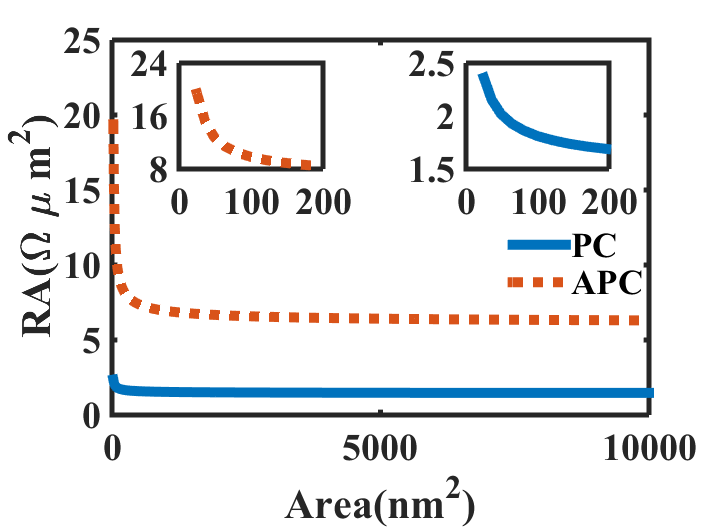}}
	
	\caption{Resistance and RA product profile of trilayer MTJ. Variation of device resistance for (a) Square cross-section and (b) Circular cross-section, for both PC and APC. Variation of RA product with cross-sectional area of trilayer MTJ, for (c) Square cross-section and (d) Circular cross-section for both PC and APC. Insets of (c) and (d) show variation of RA product for the smaller area for APC and PC respectively.}
	\label{tri_Res_RA}
\end{figure}
\indent We now discuss how RA product is affected by the area. During calculation of the areal effect, the voltage was kept constant at a value 0.01 V and from Fig. \ref{Tri_TMR_current_area} we note that the current increases with area leading to a decrease in resistance as shown in Fig. \ref{tri_Res_RA} (a) and (b). The variation of RA product for both types of geometrical structures is shown in Fig. \ref{tri_Res_RA} (c) and (d). From Fig. \ref{tri_Res_RA}, we note that the resistance is very high at smaller areas leading to a greater magnitude of the RA product as seen from Fig. \ref{tri_Res_RA}. The RA product for both PC and APC changes rapidly with the area at smaller areas due to a rapid change of resistance in that range. For both the device structures, we notice that the change in the RA product for the APC is much steeper than that of the PC due to the sharper change in resistance for the APC than the PC as can be seen from Fig. \ref{tri_Res_RA}. This sharp change in resistance in the APC is due to a rapid change in current as the area reduces as can be noted from Fig. \ref{Tri_TMR_current_area} (c) and (d). This rapid change in current can be justified by seeing the nature of the change in mode currents in APC at smaller areas as can be seen in Fig. \ref{trilayer_mode_profie} (a) and (b). The rapid change in RA product with area at the smaller scale is highlighted in the inset of Fig. \ref{tri_Res_RA} (c) and (d). As the area is increased further, device resistance decreases keeping RA product constant. At larger areas for both square and circular cross-sections, the RA product maintains a constant value of $\sim$6.3 $\Omega \mu m^2$ for APC and $\sim$1.46 $\Omega \mu m^2$ for PC. At the smallest areas considered, the RA product value for PC and APC are 2.471 $\Omega \mu m^2$ (2.361 $\Omega \mu m^2$ ) and 21.9 $\Omega \mu m^2$ (19.38 $\Omega \mu m^2$) respectively for the square (circular) cross-section. \\

\subsection{Pentalayer RTMTJ}
\begin{figure}[h]
	\subfigure[]{\includegraphics[scale=0.28]{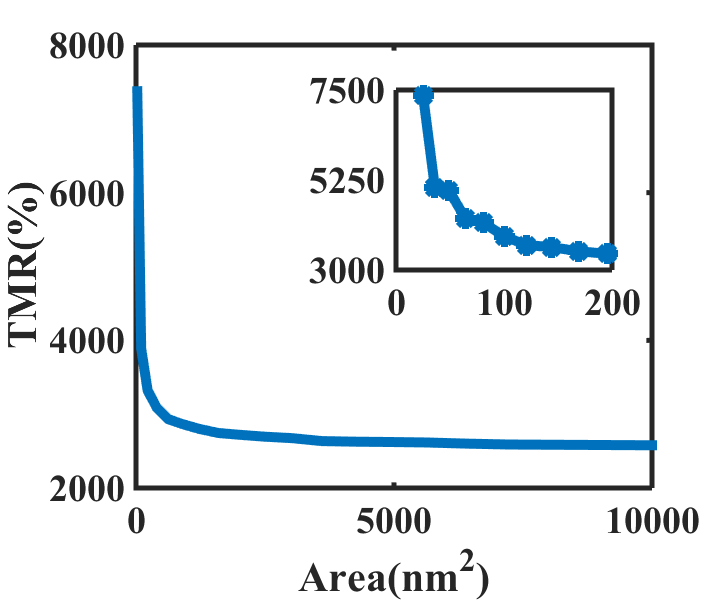}}
	\hfill
	\subfigure[]{\includegraphics[scale=0.28]{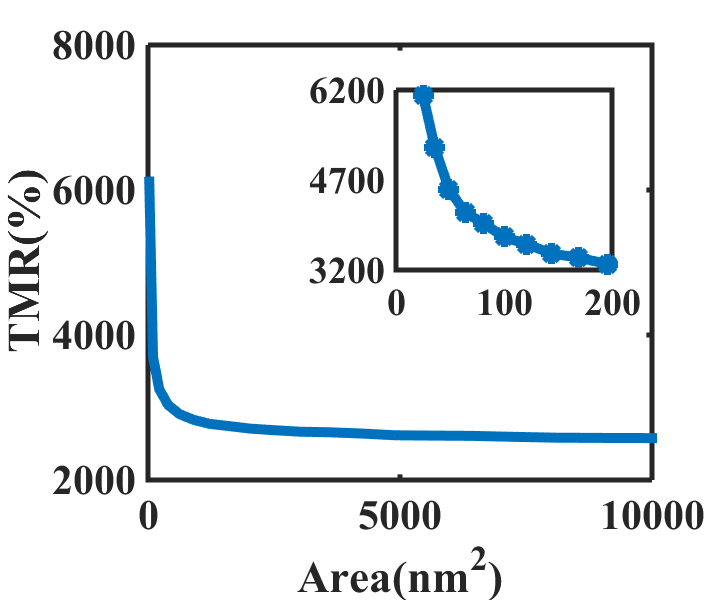}}
	\subfigure[]{\includegraphics[scale=0.28]{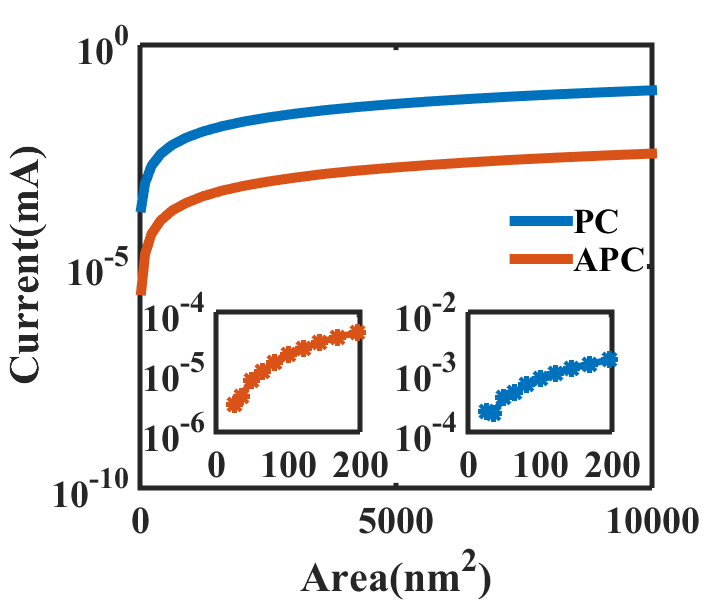}}
	\hfill
	\subfigure[]{\includegraphics[scale=0.28]{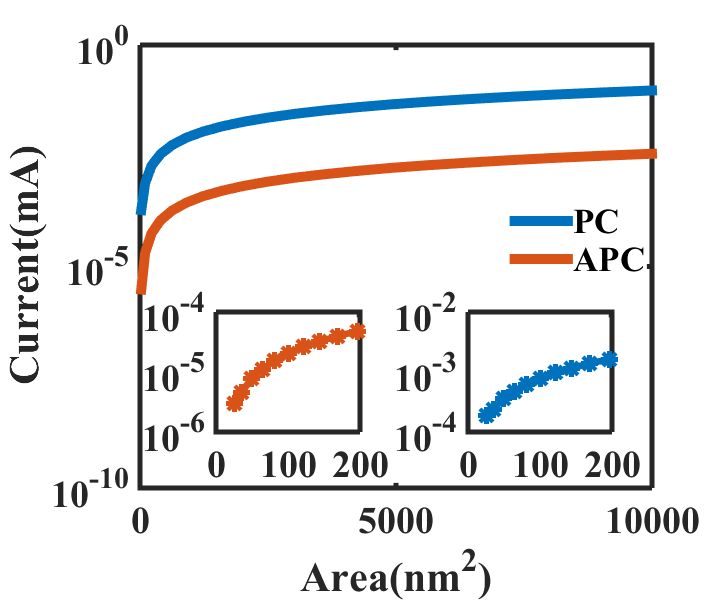}}
	\caption{TMR and total charge current profile for pentalayer RTMTJ. Variation of TMR with cross-sectional area for (a) Square cross-section and (b) Circular cross-section. Insets show variation of TMR for the smaller areas. Variations in charge currents with cross-sectional area for pentalayer RTMTJ for (c) Square cross-section and (d) Circular cross-section in PC and APC. Variation of charge current for the smaller area for APC and PC is shown in the inset of (c) and (d). The `*' symbols in the inset figures denote data points.}
	\label{Penta_TMR_current}
\end{figure}

\begin{figure}[h]
	\subfigure[]{\includegraphics[scale=0.28]{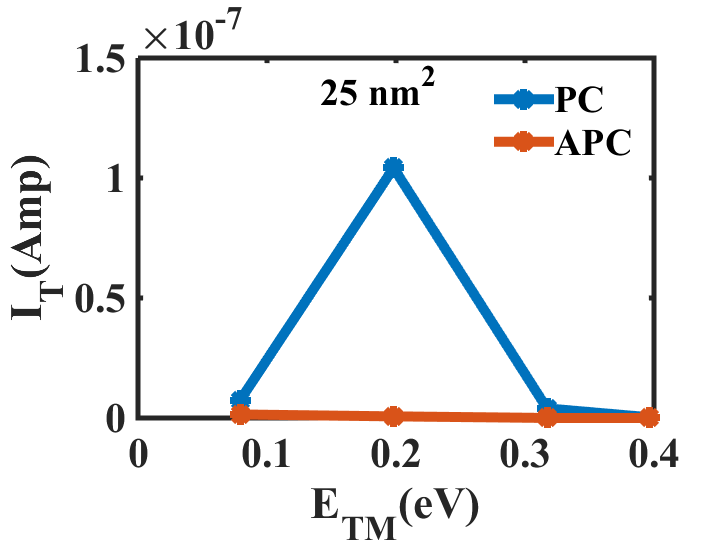}}
	\hfill\
	\subfigure[]{\includegraphics[scale=0.28]{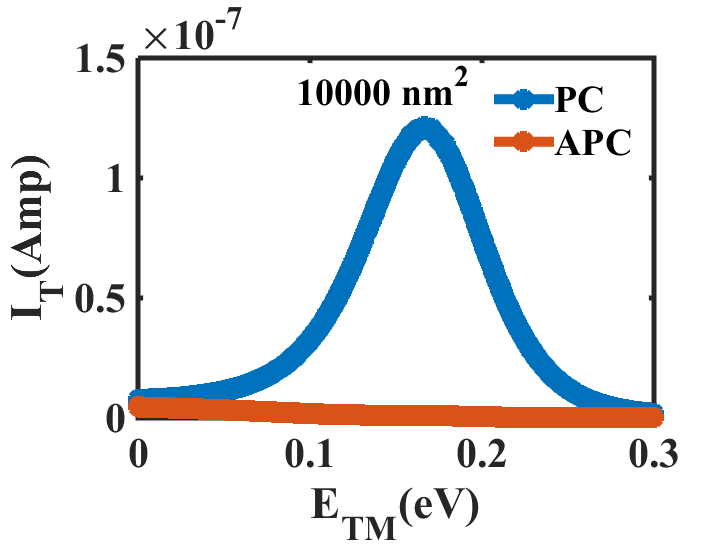}}
	\vfill
	\subfigure[]{\includegraphics[scale=0.28]{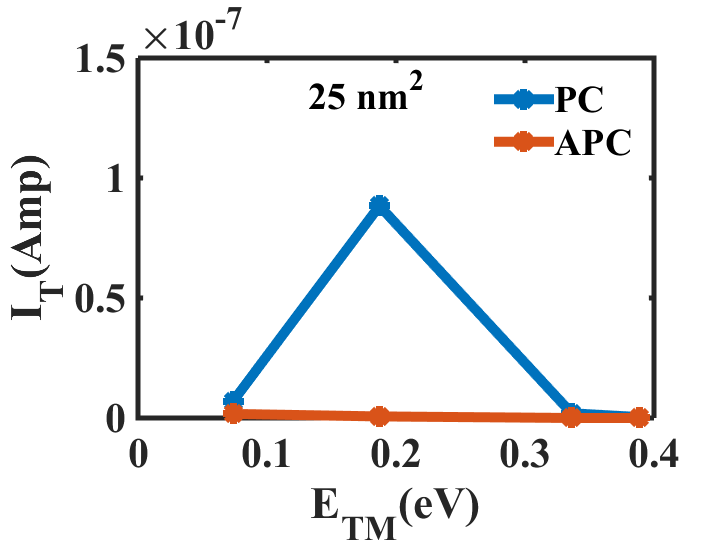}}
	\hfill
	\subfigure[]{\includegraphics[scale=0.28]{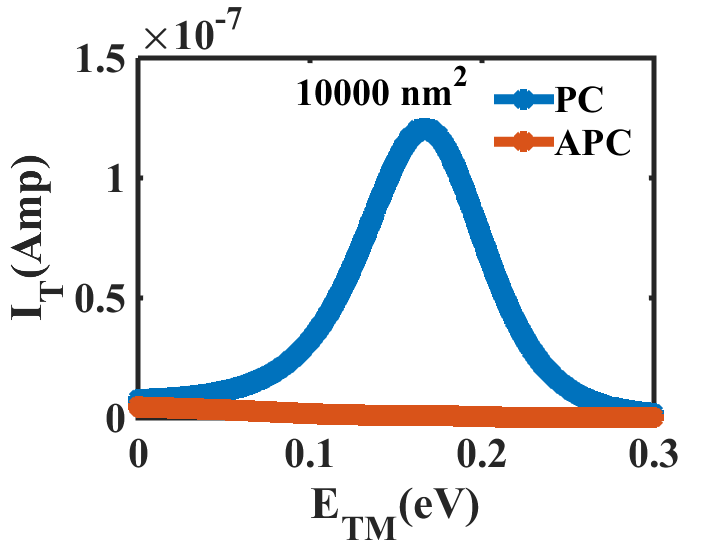}}
	\caption{Transverse mode profile for pentalayer RTMTJ devices. Variations in transverse mode currents with mode energies for square cross-sectional areas of (a) $25\ nm^2$, (b) $10000\ nm^2$ and circular cross-sectional area of (c) $25\ nm^2$ and (d) $10000\ nm^2$ in PC and APC.}
	\label{Penta_mode}
\end{figure}

In this section, we perform a similar analysis for the RTMTJ case. Starting with TMR, as shown in Fig. \ref{Penta_TMR_current}, we can see that the RTMTJ also shows constant TMR at larger areas, but increases with decreasing area and blows up rapidly at smaller areas, a trend that was noted also in the trilayer MTJ case. Although 
we notice a similar pattern like that of the trilayer MTJ, at smaller areas, unlike the trilayer MTJ, the pentalayer TMR does not show a smooth variation. For the square cross-section, at 36 $nm^2$ and 64 $nm^2$, the TMR achieves a lesser value than the expected. The reason behind this lower value of TMR becomes clear when we see the variation of charge current with area, as shown in the inset of Fig. \ref{Penta_TMR_current} (c). The variation of currents in the APC is quite smooth, whereas for the PC, it decreases at 36 $nm^2$ and 64 $nm^2$ which reduces the value of TMR than that expected. For a circular cross-section,  the TMR shows a smooth variation. For the smallest area considered (25 $nm^2$), we obtain an ultra-high TMR of 7374\% and 6116\% for the square and circular cross-section respectively. At larger areas, the TMR becomes 2579\% for both square and circular cross-sectional areas consistent with previously reported values \cite{Sharma2016}. The physics behind this ultra-high TMR can be explained from the charge current plot and the mode profile plot as shown in Fig. \ref{Penta_TMR_current} and \ref{Penta_mode} respectively. Ultrahigh TMR obtained at the smallest area is due to the huge difference between currents in the PC and APC as shown in Fig \ref{Penta_TMR_current}. We can clarify further if we analyze the mode profile of the device at that area as shown in Fig. \ref{Penta_mode}(a) and (c). It is seen that the transverse mode currents in the APC are much smaller than that in the PC, thus creating a high ratio between $I_P$ and $I_{AP}$.  In turn, it creates an ultrahigh TMR as can be seen from (\ref{TMR_eq}). The advantage of RTMTJ over MTJ is that even at larger areas we get high values of TMR, which can be explained from the mode profiles for the largest area as shown in Fig. \ref{Penta_mode}(b) and (d), for square and circular cross-sectional areas respectively. We see that the transverse mode currents are much larger for PC than APC, which is why $I_P$ is much greater than $I_{AP}$, thus resulting in a high TMR. At large areas, the modes become quasi-continuous and currents increase at a constant rate due to which the TMR remains constant at large areas.

\begin{figure}[h]
	\subfigure[]{\includegraphics[scale=0.28]{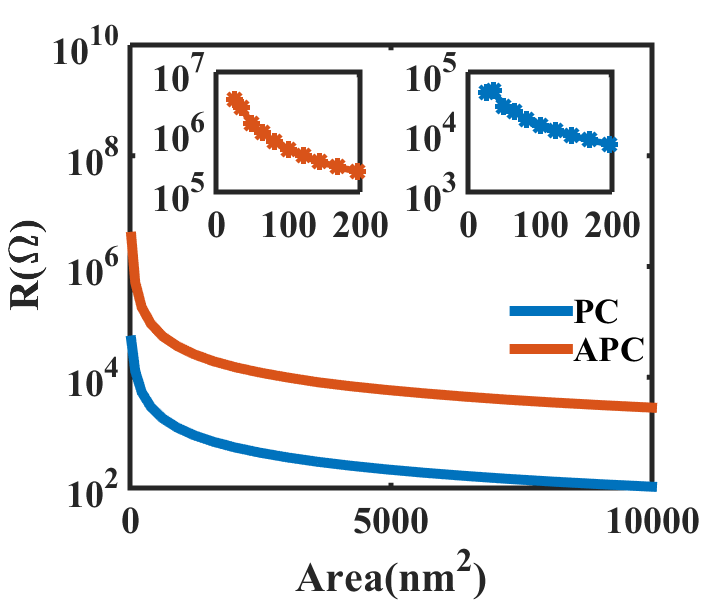}}
	\hfill
	\subfigure[]{\includegraphics[scale=0.28]{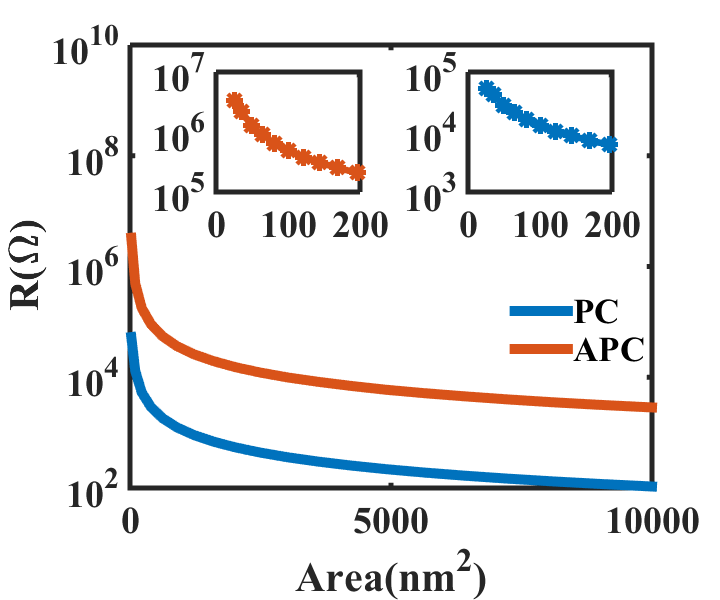}}
	
	\subfigure[]{\includegraphics[scale=0.28]{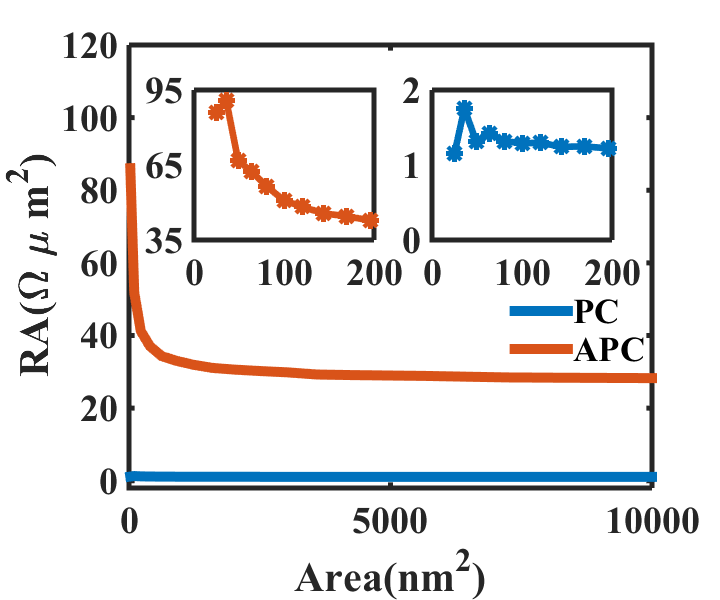}}
	\hfill
	\subfigure[]{\includegraphics[scale=0.28]{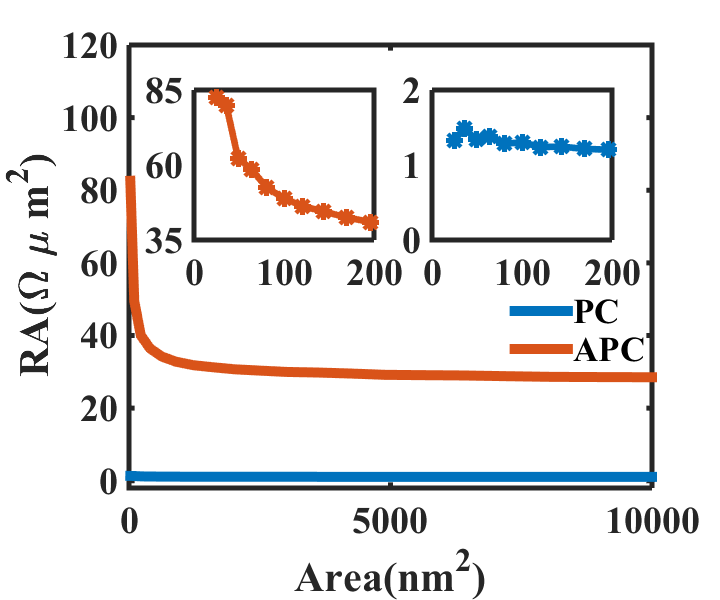}}
	\caption{Resistance and RA product profile of pentalayer RTMTJ. Variation of Resistance with cross-sectional area for (a) Square cross-section and (b) Circular cross-section in the PC and the APC. Variations at smaller areas for PC and APC are shown in the insets of (a) and (b) respectively. Variation of RA product with cross-sectional area for (c) Square cross-section and (d) Circular cross-section for the APC and the PC. Variation of RA product at smaller areas for both APC and PC is shown in the inset of (c) and (d). The `*' symbols in the inset figures denote data points.}
	\label{Penta_Res_RA}
\end{figure}
\indent Next, we describe the variation of the RA product for the RTMTJ devices with cross-sectional area. As seen from Fig. \ref{Penta_mode}, it is expected that there will be a large resistance for APC, whereas for PC it is much smaller as shown in Fig. \ref{Penta_Res_RA} (a) and (b). In the RTMTJ case, the variation of RA product with the cross-sectional area for both square and circular cross-sections look similar to those of the trilayer MTJ but the variation at smaller areas is quite different. In the APC, for a square cross-section, the RA product at 25 $nm^2$ shows a lower value than that obtained at 36 $nm^2$. In the PC both square and circular cross-sections show an oscillatory behavior of the RA product at smaller areas as shown in the inset of Fig. \ref{Penta_Res_RA} (c) and (d), where the oscillation is more visible in the square cross-section case. In the PC due to the resonant condition at 25 $nm^2$ and 49 $nm^2$, the RTMTJ with square cross-section allows more current to flow which is shown in the inset of Fig. \ref{Penta_TMR_current} (c) causing the resistance in PC to be low at those areas which can be seen in Fig. \ref{Penta_Res_RA} (c). This causes the RA product to be small for those areas. At the smallest area considered, the RA product is 86.08 $\Omega\ \mu m^2$ and 82.6 $\Omega\ \mu m^2$ for the square and circular cross-sections respectively. Although at the beginning, at 36 $nm^2$, the RA product increases for square cross-section only, it starts to decrease as the area is increased and saturates at a value of 28 $\Omega\ \mu m^2$ for larger areas for both square and circular cross-sections. Although in PC the RA product oscillates at smaller areas, at the larger areas it maintains a constant value $\sim$1 $\Omega\ \mu m^2$ for both structures as shown in Fig. \ref{Penta_Res_RA} (c) and (d).

\subsection{Switching Characteristics}
\indent In the previous section, we described the impact of scaling on various properties of MTJ and RTMTJ which are related to the read processes. Now, we will describe how scaling affects the writing scheme in these devices. The magnetization switching process in an MTJ with perpendicular anisotropy is dominated either by the macrospin assumption or by subvolume excitation, depending on the cross-sectional area of the MTJ. In an experiment by Sun et. al.\cite{sun2011effect}, it was shown that at larger cross-sectional areas ($\sim$50 nm in lateral size), the switching is dominated by subvolume excitation, whereas for smaller areas it depends mainly on the macrospin-like process which was also shown by Ohuchida et. al., \cite{ohuchida2015impact} using the LLG simulation. In the macrospin theory, it is assumed that the free layer is homogeneously magnetized so that a single spin entity is sufficient to describe the magnetization dynamics. Subvolume excitation is described as that which occurs due to a smaller anisotropy field at the center of the free layer than the edges, which generates a subvolume domain\cite{Sun2013,ito2014dependence} that is pushed toward the edges during magnetization reversal of the free layer. This ends up creating a domain wall in the free layer. Sun et. al., \cite{sun2011effect} showed that for a device of size 120$\times$120 $nm^2$,  the subvolume size is of the order of$\sim$40 nm. As we are investigating the scaling effect and as established in the previous section that all the device characteristics vary only at regions with smaller cross-sectional areas, we focus only on the smaller areas ranging from 25-400 $nm^2$  for studying the switching characteristics of MTJ and RTMTJ. In this range of areas, we can thus safely neglect the subvolume excitation. As we are dealing with magnets with perpendicular anisotropy, the critical spin currents for switching 
is given by
\begin{equation}
I_{s_c}=\frac{2q\alpha}{\hbar} M_S V H_{K_{eff}}\label{Isc}\\
\end{equation} 
where $H_{K_{eff}}=H_{K_{\perp}}-H_d$ is the effective anisotropy field. Here $H_{K_{\perp}}$ is the perpendicular anisotropy field and $H_d$ is the demagnetization field. Here V is the volume of the free magnet. For investigating the switching characteristics, we considered a free layer thickness of 1 nm and a saturation magnetization $M_S$=1100 emu/cc. For any magnet, retention time is given by N\.{e}el-Arrhenius equation $\tau=\tau_0\ exp\left( \frac{K_uV}{k_BT}\right)$. Here $K_uV$ represents the energy barrier, separating two stable states along a uniaxial direction and $1/\tau_0$($\sim$ 1 GHz) is called attempt frequency. $K_B$ is the Boltzmann constant and $T$ is the temperature. Starting with the smallest area considered and assuming an energy barrier of $40k_BT$, we find in the trilayer MTJ case, that the AP-P critical switching voltage is $\sim$ 1 volt, whereas due to insufficient Slonczewski-torque P-AP switching is not possible. To keep the energy barrier of $40k_BT$ in the free layer FM, an effective anisotropy field of $H_{K_{eff}}$=$1.16\times 10^5$ Oe is estimated, which is quite impossible to achieve. To overcome this problem and to study the effect of scaling, we lower the value of $H_{K_{eff}}$ to $6.95\times 10^3$ Oe, compromising on the thermal stability for lower cross-sectional areas $<400\ nm^2$ at room temperature. For practical application of these ultra-small-scale devices($<400\ nm^2$) for memory application either we need to find a material with very high effective anisotropy field or these devices can still be used with our assumed parameters but should be operated at very low temperature for a longer data retention time of $\sim$10 years. In our device, we calculated the operating temperature($T$) should be below 17.31$K$, for the cross-sectional area of 25 $nm^2$, whereas $T$ increases with the cross-sectional area. Although an in-depth calculation\cite{Sun2013,zhang2015compact} reveals that at smaller areas the energy barrier is enhanced which can increase the operating temperature of our device.
\begin{figure}[h]
	\subfigure[]{\includegraphics[scale=0.28]{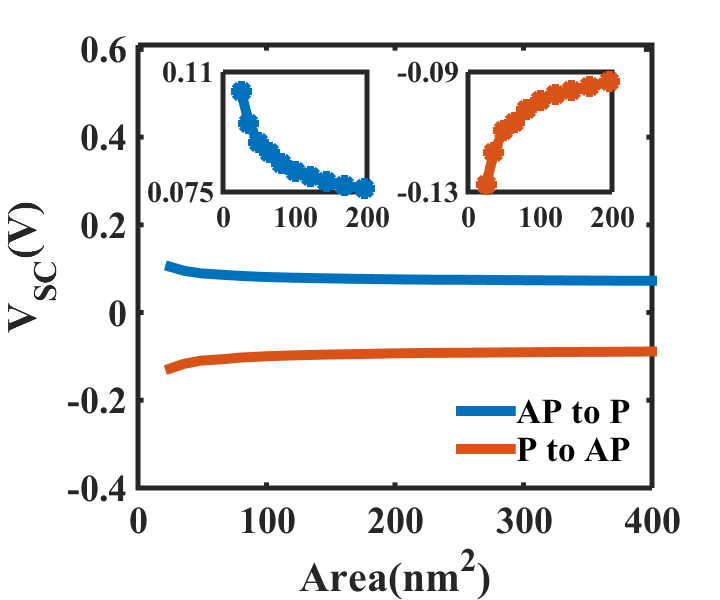}}
	\hfill
	\subfigure[]{\includegraphics[scale=0.28]{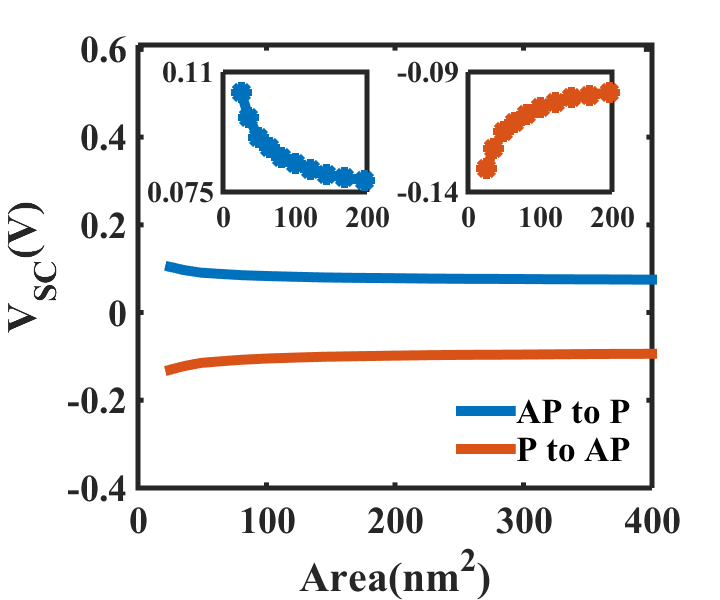}}
	\caption{Variation of critical switching voltage($V_{SC}$) for AP-P and P-AP switching with cross-sectional area in the case of trilayer MTJ, for (a) Square cross-section and (b) Circular cross-section. Critical switching voltage profile remains constant at higher areas, whereas it shows a rapid change as the cross-sectional area is decreased at an ultra-small-scale. In both figures change in $V_{SC}$ with cross-sectional area at this ultra-small-scale is shown for switching from AP-P in left inset and from P-AP in right inset. `*' symbols in the inset figures denote data points.}
	\label{tri-switch}
\end{figure} 
 
\begin{figure}[h]
	\subfigure[]{\includegraphics[scale=0.28]{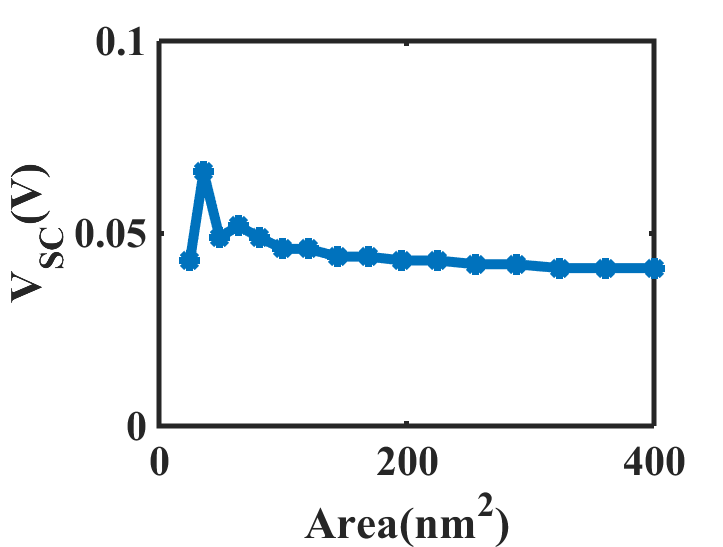}}
	\hfill
	\subfigure[]{\includegraphics[scale=0.28]{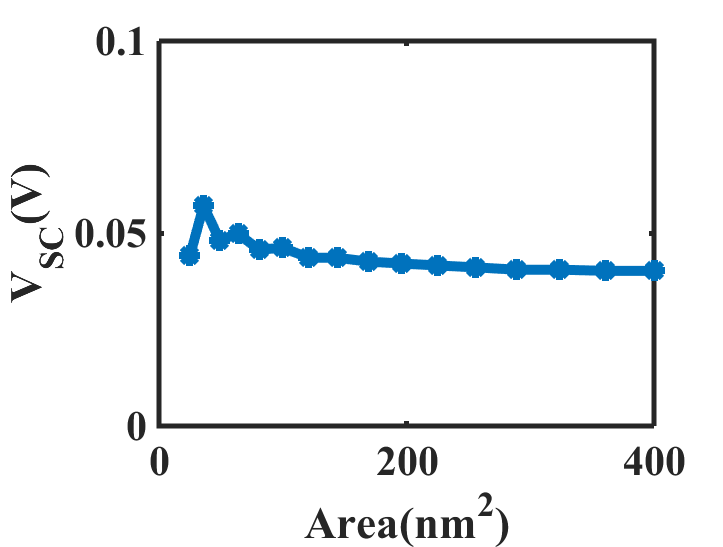}}
	\caption{Variation of critical switching voltage for AP-P switching with cross-sectional area in the case of pentalayer RTMTJ for (a) Square cross-section and (b) Circular cross-section. Due to resonant condition at certain smaller area allows more current to flow which switches the magnet at lower value of voltage. }
\label{Penta_switch}
\end{figure}
\indent First, we calculate the critical switching voltage ($V_{sc}$) of trilayer MTJ for both square and circular cross-section as shown in Fig \ref{tri-switch}. For both structures, we see the switching voltage is higher at a smaller area and becomes constant as the area is increased. This is simply because, at smaller areas, discrete modes with large energy spacings are present as shown in Fig. \ref{trilayer_mode_profie} (a) and (e), which results in lower currents. Thus to attain critical switching currents one needs to apply larger voltages. As the cross-sectional area is increased, the critical switching current is increased linearly with the area(volume), as can be seen from (\ref{Isc}). At larger areas, where transverse modes become quasi-continuous, currents and thus spin currents increase linearly with area. Although critical spin current increases with increasing area, at higher areas due to quasi-continuous conducting modes, spin currents also increase in a similar fashion, thus maintaining a constant critical switching voltage for both configurations. Our simulations show that the critical switching voltage from antiparallel to parallel(AP-P) and parallel to antiparallel (P -AP) switching at the smallest area are 0.1044 V(0.1041 V) and -0.128 V(-0.1304 V) respectively, for the square(circular) cross-section, whereas at larger areas they are 0.0666 V(0.0695 V) and -0.0816 V(-0.0868 V) respectively.\\
\indent  In the RTMTJ case, it is found that P-AP switching cannot be accomplished due to insufficient Slonczewski-like spin torque. So, we discuss the AP-P switching, shown in Fig \ref{Penta_switch}, where we see at that at smaller areas $V_{SC}$ is oscillatory. From the concept of increasing number of modes with the area, it is expected to have a switching profile similar to the MTJ, but due to the resonance situation, certain cross-sections (25 $nm^2$,49 $nm^2$,81 $nm^2$) allow more current to flow which helps to switch at comparatively lower voltages. The highest $V_{SC}$ we obtained is 0.066 V and 0.057 V for the square and circular cross-sections respectively with an area of 36 $nm^2$. At 400 $nm^2$, this value is 0.041 V and 0.040 V for the square and circular cross-sections respectively.

\section{Device Proposal under Scaling}
\begin{figure}[h]
	\subfigure[]{\includegraphics[scale=0.2]{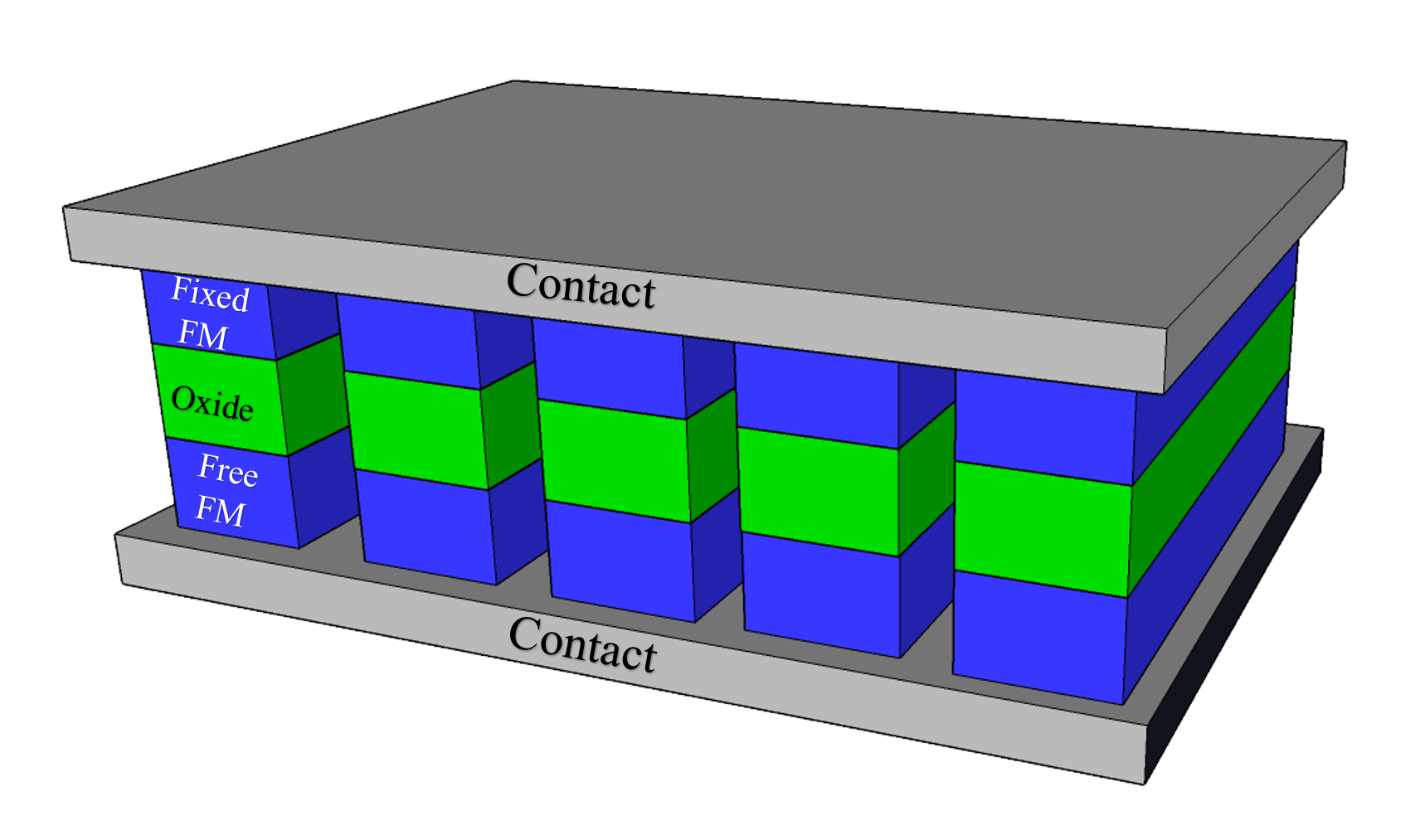}}
	\hfill
	\subfigure[]{\includegraphics[scale=0.2]{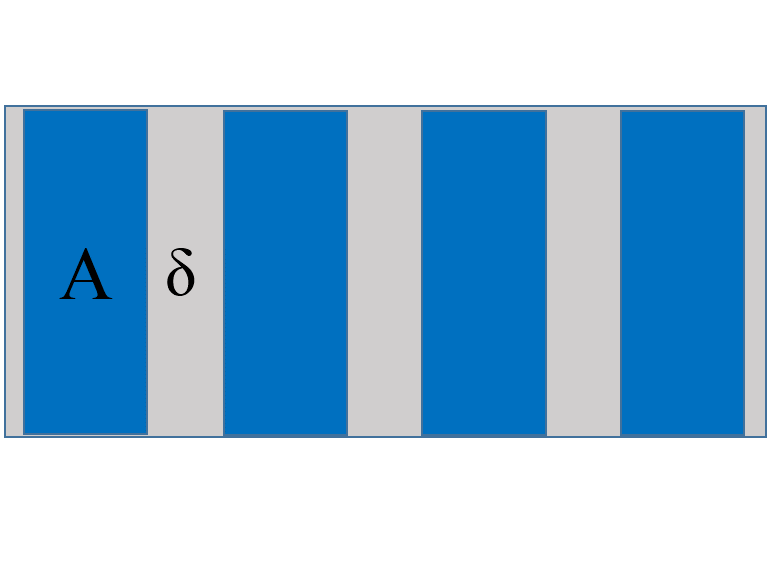}}
	
	\caption{Schematic of our proposed device, which has identical MTJs with small cross-sectional area connected to the contacts in parallel to each other. (a) Schematic 3D view of our proposed device. (b) A cross-sectional view of the junction between ferromagnet and contact. Blue region represents cross-section of FM layer with an area `A' and gray region represents the area between two consecutive MTJ with an area `$\delta$'.
	\label{Dev_Schem}
		}	
\end{figure}
The devices considered here, i.e., trilayer MTJ, and pentalayer RTMTJ, show high TMR, high resistance and low RA product at smaller areas. This high TMR and low RA product satisfy our application requirements, but high resistance can limit our device applications. We propose a device configuration that comprises MTJs with a smaller area connected with each other in a parallel configuration so that every MTJ has a common contact at both sides as shown in Fig. \ref{Dev_Schem}. The overall resistance of the device in a particular configuration will be R/n if n such MTJs are connected in parallel which is much smaller if n is large. If the individual resistance of the MTJ in PC and APC are $R_P$ and $R_{AP}$ respectively, then total resistance of our proposed device would be $R_P/n$ and $R_{AP}/n$ in the respective configurations. Thus from Eq. (\ref{TMR_eq}), it is seen that there would be no effect on TMR. If the cross-sectional area of each MTJ is $A$ and the area between two consecutive MTJs is $\delta$ then the RA product of our proposed device would be
\begin{align}
RA_{new} &=\frac{R}{n}\left[ n A + (n-1)\delta\right]\nonumber\\
 &=R\left[A+(1-\frac{1}{n})\delta\right]. 
\end{align}
If n is large and $\delta$ is small, $RA_{new}\approx RA$ which is the same as that of a single MTJ cell with a small area. The advantage of our proposed device over a single MTJ with an equivalent area will be much more clear if we compare the TMR of both devices. If n is large then the overall cross-section of our proposed device is much larger. In this situation, the resistance of our device is much lesser than the resistance of a single MTJ with smaller cross-sectional areas. It also maintains a high TMR value obtained in the single MTJ with small cross-section and if $\delta$ is small it also maintains the RA product at the same value at smaller areas. On the other hand, although a single MTJ with a large equivalent area has a low resistance and a low RA product as seen from the characteristics we have obtained from the simulation, its TMR value would be much less than our proposed device. This is because we have seen that at large areas, the TMR becomes much less than that obtained at a small areas for both MTJ and RTMTJ. This shows that for application purposes, our proposed device is a better candidate than the conventional single MTJ.\\ Our proposed device can also be used as a spin torque nano-oscillator(STNO), where a high TMR, low RA product, and a low resistance are together required. In this configuration, the magnetization of the free layer can be perturbed by the dipolar field of adjacent MTJ cells, which will affect the switching voltage. 

\section{Conclusion}
We investigated the scaling of magnetic tunnel junction devices in the trilayer and pentalayer configurations by varying the cross-sectional areas along the transverse direction using nonequilibrium Green's function spin transport formalism. We studied the geometry dependence by considering both square and circular cross-sections. As the transverse dimension in each case reduces, we demonstrated that the transverse mode energy profile plays a major role in the resistance-area product. Both types of devices showed constant tunnel magnetoresistance at larger cross-sectional areas but achieved ultra-high magnetoresistance at small cross-sectional areas, while maintaining low resistance-area products. We noticed that although the critical switching voltage for switching the magnetization of the free layer nanomagnet in the trilayer case remained constant at larger areas, it needs more energy to switch at smaller areas. In the pentalayer case, we observed an oscillatory behavior at smaller areas as a result of double barrier tunneling. We also described how switching characteristics of both kinds of devices are affected by the scaling. We also proposed a device structure that utilizes the advantage of high TMR at smaller areas and hence reduces the total resistance.
\section{Acknowledgement}
This work was supported partially by IIT Bombay SEED grant. D. Das would like to thank A. Sharma for useful discussions.
\bibliographystyle{IEEEtran}
\bibliography{Reference}

\end{document}